\documentclass[twocolumn,prl,superscriptaddress,a4paper,showpacs]{revtex4}

\usepackage{amsmath,amssymb}
\usepackage{epsfig}
\usepackage{citesort}
\usepackage{epic,eepic}

\newcommand{\bs}{\ensuremath{\mathbf{s}}}
\newcommand{\bt}{\ensuremath{\mathbf{t}}}

\newcommand{\nbk}{\ensuremath{n^{(k)}}}

\newcommand{\ZZ}{\ensuremath{\mathbb{Z}^2}}

\newcommand{\dnv}{
\setlength{\unitlength}{0.07em}
\begin{picture}(8,10)(-4,3)
\path(0,0)(0,10)
\put(0,0){\whiten\circle{4}}
\put(0,10){\whiten\circle{4}}
\end{picture}}

\newcommand{\dnh}{
\setlength{\unitlength}{0.07em}
\begin{picture}(18,6)(0,-2)
\path(4,0)(14,0)
\put(4,0){\whiten\circle{4}}
\put(14,0){\whiten\circle{4}}
\end{picture}}

\newcommand{\dnp}{
\setlength{\unitlength}{0.07em}
\begin{picture}(8,6)(0,-2)
\put(4,0){\whiten\circle{4}}
\end{picture}}

\newcommand{\dnsq}{
\setlength{\unitlength}{0.07em}
\begin{picture}(18,10)(-4,2)
\path(0,0)(10,0)
\path(10,0)(10,10)
\path(0,10)(10,10)
\path(0,0)(0,10)
\put(0,0){\whiten\circle{4}}
\put(10,0){\whiten\circle{4}}
\put(0,10){\whiten\circle{4}}
\put(10,10){\whiten\circle{4}}
\end{picture}}

\newcommand{\dnvdp}[2]{
\setlength{\unitlength}{0.07em}
\begin{picture}(10,10)(-4,3)
\path(0,0)(0,10)
\put(0,0){\circle*{4}}
\put(3,-4){{\tiny #1}}
\put(0,10){\circle*{4}}
\put(3,10){{\tiny #2}}
\end{picture}}

\newcommand{\dnhdp}[2]{
\setlength{\unitlength}{0.07em}
\begin{picture}(21,6)(-2,-2)
\path(4,0)(14,0)
\put(4,0){\circle*{4}}
\put(-2,-4){{\tiny #1}}
\put(14,0){\circle*{4}}
\put(16,-4){{\tiny #2}}
\end{picture}}

\newcommand{\dnpdp}[1]{
\setlength{\unitlength}{0.07em}
\begin{picture}(10,6)(0,-2)
\put(4,0){\circle*{4}}
\put(7,-4){{\tiny #1}}
\end{picture}}

\newcommand{\dnsqdp}[4]{
\setlength{\unitlength}{0.07em}
\begin{picture}(22,10)(-6,2)
\path(0,0)(10,0)
\path(10,0)(10,10)
\path(0,10)(10,10)
\path(0,0)(0,10)
\put(0,0){\circle*{4}}
\put(-6,-4){{\tiny #1}}
\put(10,0){\circle*{4}}
\put(13,-4){{\tiny #2}}
\put(0,10){\circle*{4}}
\put(-6,10){{\tiny #3}}
\put(10,10){\circle*{4}}
\put(13,10){{\tiny #4}}
\end{picture}}

\newcommand{\dntp}{
\setlength{\unitlength}{0.07em}
\begin{picture}(18,10)(-4,2)
\path(0,0)(10,0)(5,8.7)(0,0)
\put(0,0){\whiten\circle{4}}
\put(10,0){\whiten\circle{4}}
\put(5,8.7){\whiten\circle{4}}
\end{picture}}

\newcommand{\dntm}{
\setlength{\unitlength}{0.07em}
\begin{picture}(18,10)(-4,2)
\path(0,8.7)(10,8.7)(5,0)(0,8.7)
\put(10,8.7){\whiten\circle{4}}
\put(0,8.7){\whiten\circle{4}}
\put(5,0){\whiten\circle{4}}
\end{picture}}

\newcommand{\dntr}{
\setlength{\unitlength}{0.07em}
\begin{picture}(14,10)(-4,2)
\path(0,0)(5,8.7)
\put(0,0){\whiten\circle{4}}
\put(5,8.7){\whiten\circle{4}}
\end{picture}}

\newcommand{\dntl}{
\setlength{\unitlength}{0.07em}
\begin{picture}(14,10)(0,2)
\path(10,0)(5,8.7)
\put(10,0){\whiten\circle{4}}
\put(5,8.7){\whiten\circle{4}}
\end{picture}}

\newcommand{\dntpdp}[3]{
\setlength{\unitlength}{0.07em}
\begin{picture}(22,10)(-6,2)
\path(0,0)(10,0)(5,8.7)(0,0)
\put(0,0){\circle*{4}}
\put(-6,-4){{\tiny #1}}
\put(10,0){\circle*{4}}
\put(12,-4){{\tiny #2}}
\put(5,8.7){\circle*{4}}
\put(8,8){{\tiny #3}}
\end{picture}}

\newcommand{\dntrdp}[2]{
\setlength{\unitlength}{0.07em}
\begin{picture}(18,10)(-6,2)
\path(0,0)(5,8.7)
\put(0,0){\circle*{4}}
\put(-6,-4){{\tiny #1}}
\put(5,8.7){\circle*{4}}
\put(8,8){{\tiny #2}}
\end{picture}}

\newcommand{\dntldp}[2]{
\setlength{\unitlength}{0.07em}
\begin{picture}(18,10)(0,2)
\path(10,0)(5,8.7)
\put(10,0){\circle*{4}}
\put(12,-4){{\tiny #1}}
\put(5,8.7){\circle*{4}}
\put(8,8){{\tiny #2}}
\end{picture}}

\newcommand{\dntmaxr}{
\setlength{\unitlength}{0.07em}
\begin{picture}(23,10)(-4,2)
\path(0,0)(10,0)(5,8.7)(0,0)
\path(5,8.7)(15,8.7)(10,0)
\put(0,0){\whiten\circle{4}}
\put(10,0){\whiten\circle{4}}
\put(5,8.7){\whiten\circle{4}}
\put(15,8.7){\whiten\circle{4}}
\end{picture}}

\newcommand{\dntmaxl}{
\setlength{\unitlength}{0.07em}
\begin{picture}(22,10)(-9,2)
\path(0,0)(10,0)(5,8.7)(0,0)
\path(5,8.7)(-5,8.7)(0,0)
\put(0,0){\whiten\circle{4}}
\put(10,0){\whiten\circle{4}}
\put(5,8.7){\whiten\circle{4}}
\put(-5,8.7){\whiten\circle{4}}
\end{picture}}

\newcommand{\dntmaxup}{
\setlength{\unitlength}{0.07em}
\begin{picture}(18,14)(-4,-4)
\path(0,0.7)(10,0.7)(5,-8)(0,0.7)
\path(0,0.7)(5,9.4)(10,0.7)
\put(10,0.7){\whiten\circle{4}}
\put(0,0.7){\whiten\circle{4}}
\put(5,-8){\whiten\circle{4}}
\put(5,9.4){\whiten\circle{4}}
\end{picture}}

\begin{document}

\title{A density functional theory for general hard-core lattice gases}

\date{\today}

\author{Luis Lafuente}
\email{llafuent@math.uc3m.es}

\author{Jos\'e A.~Cuesta}
\email{cuesta@math.uc3m.es}

\affiliation{Grupo Interdisciplinar de Sistemas Complejos (GISC),
Departamento de Matem\'aticas, Universidad Carlos III de Madrid,
Avenida de la Universidad 30, E-28911, Legan\'es, Madrid, Spain}

\begin{abstract}
We put forward a general procedure to obtain an approximate
free energy density functional
for any hard-core lattice gas, regardless of the shape of the particles,
the underlying lattice or the dimension of the system. The procedure is 
conceptually very simple and recovers effortlessly previous results for
some particular systems. Also, the obtained density functionals belong to 
the class of fundamental measure functionals and, therefore, are always 
consistent through dimensional reduction. We discuss possible extensions
of this method to account for attractive lattice models.
\end{abstract}


\maketitle

%One cannot conceive the development of Statistical Physics
%without lattice models. On the one hand,
%their simplicity allows for treatments which would be 
%infeasible for continuum models. On the other hand, there
%is virtually no phenomenon (either classical or quantum)
%which cannot be observed in some suitable lattice model. 
%%And there
%%is yet another reason justifying the crucial role of lattice
%%models: universality, i.e.\ the fact that near critical points
%%the details of the interaction are only marginally relevant,
%%and totally different models (both lattice or continuum models)
%%may exhibit the same behavior. It is thus
%Not surprisingly, most paradigms of Statistical Physics
%are lattice models.
%%, and that their study experienced a ``golden age''
%%during the development of the theory of phase transitions and
%%critical phenomena \cite{lavis:1999}.
%
Despite %this, 
the crucial role that lattice models have had in the development
of Statistical Physics, when one looks for such models in the
literature of density functional theory, the results are
scarce. In the last few years, though, 
%there have been several
%contributions filling this gap, and 
some of the most classical approximations
%(Ramakrishnan-Yussouff, weighted-density
%approximation, etc.; see ref.~\cite{evans:1992} for a review
%of these theories)
have been extended to lattice systems
\cite{nieswand,reinhard:2000,prestipino:2003} and used to
study different phenomena (like freezing and fluid-solid
interfaces \cite{nieswand,prestipino:2003} or confined fluids
\cite{reinhard:2000}).
Also very recently, fundamental measure (FM) theory
has been added to the list through
its formulation for systems of parallel
hard hypercubes in hypercubic lattices \cite{lafuente:2002a}.
The construction mimics that of its continuum counterpart
\cite{cuesta:1997a}, and like the FM functional for hard
spheres, it is obtained from a zero-dimensional ($0d$) functional
(a functional for cavities holding one particle at most).
This theory possesses a remarkable property: dimensional crossover,
which allows obtaining the functional for $d-1$ dimensions
from the one for $d$ dimensions by confining the system through
an external field to lie in a $(d-1)$-dimensional slit.
Dimensional crossover has been applied to the already
mentioned system of parallel hard hypercubes in order to
obtain FM functionals for
nearest-neighbor exclusion lattice gases in two-dimensional
(square and triangular) and three-dimensional (simple,
body-centered and face-centered cubic) lattices \cite{lafuente:2003p}.

This increasing interest in density functionals for lattice models
has several motivations. On the one hand, some systems are
particularly difficult to study using continuum models. For
them, lattice models provide convenient simplifications. This
is the case of glasses \cite{glasses} or fluids in porous media
\cite{kierlik:2001}, to name only two. On the other hand, lattice 
models cover a wider range of problems, many of which do
not even belong to the theory of fluids (like roughening \cite{roughening} or
DNA denaturation \cite{dna}, to name only two) and so have never been
studied with density functional theory. Finally, from a 
purely theoretical point of view, these extensions are also
interesting because they reveal features of the
structure of the approximate functionals which are hidden or
at least not apparent in their continuum counterparts
(this is the case of FM functionals).

In this letter we propose a simple systematic procedure
to construct a FM functional for {\em any} hard-core lattice
model. The construction is based on the dimensional crossover
of this theory, much like the latest versions for continuum
models. 
%Although in order to be clear the presentation will
%be somewhat heuristic, the theory can be rigorously founded, as
%will be pointed out later.

Let us begin by realizing that all FM functionals for
lattice models studied in Refs.~\cite{lafuente:2002a,lafuente:2003p}
share a common pattern,
namely, the excess free energy can be written as
\begin{equation}
\label{eq:1}
\beta \mathcal{F}^{\text{ex}}[\rho]=\sum_{\bs\in\mathcal{L}}
\sum_{k\in\mathcal{I}} a_k \Phi_0\left(\nbk(\bs)\right),
\end{equation}
where $\mathcal{L}$ denotes the lattice,
$\mathcal{I}$ is a set of indices suitably chosen to denote the
different weighted densities $\nbk(\bs)\equiv
\sum_{\bt\in\mathcal{C}_k(\bs)} \rho(\bt)$, $a_k$ are integer
coefficients which depend on the specific model,
$\Phi_0(\eta)=\eta+(1-\eta)\ln (1-\eta)$ is the
excess free energy of a $0d$ cavity with
average occupancy $0 \leq \eta \leq 1$,
$\rho(\bs)$ is the density profile of the
system (specifically, the occupancy probability of node $\bs$)
and $\mathcal{C}_k(\bs)$ is, for each $k\in\mathcal{I}$,
a finite labeled subgraph of the lattice placed at node $\bs$
(vertices are labeled with node vectors). The shape
of the graphs $\mathcal{C}_k(\bs)$ also depends on the model.
{}From the definition, $\nbk(\bs)$ appears as the
mean occupancy of the lattice region defined by
$\mathcal{C}_k(\bs)$.

For the sake of clarity we will illustrate this formal setup and
the arguments to come
with a simple example: the two-dimensional square lattice gas
with first and second neighbor exclusion. With the help of
a diagrammatic notation already introduced in \cite{lafuente:2003b},
the excess free-energy functional for this model 
takes the form
\begin{equation}
\label{eq:2}
\beta \mathcal{F}^{\text{ex}}[\rho]=\sum_{\bs\in\ZZ}\left[
\Phi_0\left(\dnsq\right)-\Phi_0\left(\dnh\right)-
\Phi_0\left(\dnv\right)+\Phi_0\left(\dnp\right)\right],
\end{equation}
where the diagrams represent the (four in this case)
weighted densities $\dnsq=n^{(1,1)}(\bs)$, $\dnh = n^{(1,0)}(\bs)$,
$\dnv = n^{(0,1)}(\bs)$ and $\dnp = n^{(0,0)}(\bs)$, where
[$\bs=(s_1,s_2)$]
\begin{equation}
n^{(k_1,k_2)}(\bs)=\sum_{i=0}^{k_1}\sum_{j=0}^{k_2}
\rho(s_1+i,s_2+j).
\end{equation}
This notation uses explicitly the shape of the graphs
$\mathcal{C}_k(\bs)$. Thanks to this more visual representation
it is easily verified that all these graphs represent
$0d$ cavities of the lattice, because we can place at most
one particle in any of them. This is a general feature of all
FM functionals described by the pattern (\ref{eq:1}), so from
now on, the $\mathcal{C}_k(\bs)$ will be referred to as 
$0d$ cavities.

As we will make clear immediately, the form
(\ref{eq:1}) with the $\mathcal{C}_k(\bs)$
given by $0d$ cavities is a direct consequence of
the exact dimensional crossover to any $0d$ cavity that FM functionals
possess. The latter means that if we take a density
profile which vanishes outside a given $0d$ cavity (henceforth
a $0d$ profile) and evaluate the functional, we will obtain the
exact value of the free energy. The only known approximate density
functionals having this property are FM ones
\cite{cuesta:1997a,fmt,lafuente:2002a,lafuente:2003p}. (As a matter of fact,
the property can be regarded as the very constructive principle 
of FM theory\cite{fmt}.) 

Before we start let us define a
\emph{maximal cavity} to be any $0d$ cavity
which, enlarged by any lattice site, stops being a $0d$ cavity
because it can accommodate more than one particle.
Clearly, any $0d$ cavity must be contained in a maximal cavity,
so $0d$ dimensional crossover needs only be proved for maximal
cavities. (Notice that there
can be more than one maximal cavity in a given system.)
%All lattice FM functionals previously studied \cite{lafuente:2002a,
%lafuente:2003p} have been shown to have
%exact dimensional crossover to arbitrary $0d$ cavities.

Let us now try to construct the simplest possible functional
of the class (\ref{eq:1}) which fulfills the exact dimensional
crossover requirement. Its construction will 
proceed iteratively. In the first place, if
the functional must return the exact free energy
when evaluated at any $0d$ profile, there must 
appear a term in (\ref{eq:1}) for each maximal cavity
of the model, and the corresponding coefficient $a_k$ must be 1.
For the running example we are considering,
Eq.~(\ref{eq:2}), this means that we
should start off with the ansatz
\begin{equation}
\label{eq:app1}
\beta\mathcal{F}^{\text{ex}}_{1}[\rho]=\sum_{\bs\in\ZZ}\Phi_0\left(
\dnsq\right),
\end{equation}
because $\dnsq$ is the only maximal cavity of this model.
%All lattice FM functionals considered thus far \cite{lafuente:2002a,
%lafuente:2003p} contain terms corresponding to all maximal
%cavities, and all of them appear with coefficient 1.

\begin{figure}
\epsfig{file=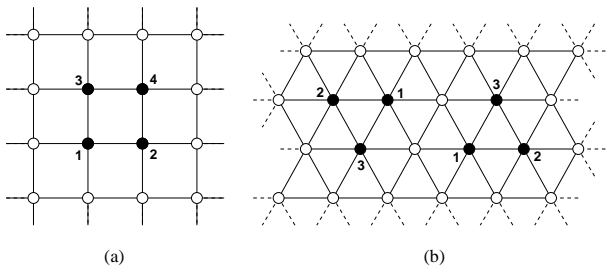,width=80mm,clip=}
\caption{Examples of $0d$ profiles corresponding to maximal cavities
for the first and second neighbor exclusion lattice gas in the
square lattice (a) and the nearest
neighbor exclusion lattice gas in the triangular lattice (b).
The density can only have nonzero values at the black nodes
(which define maximal cavities)).}
\label{fig:1}
\end{figure}

If (\ref{eq:app1}) were the final functional, evaluated at any
maximal $0d$ profile (one corresponding to a maximal cavity) it
should return the exact free energy. In the example, all maximal
$0d$ profiles have the form illustrated in Fig.~\ref{fig:1}a.
Let us now substitute this profile in (\ref{eq:app1}) and see
what comes out. For an easy way to do the evaluation, just 
imagine the graphs $\mathcal{C}_k(\bs)$ as windows which only
allow to see the content of the lattice nodes they overlap.
Then the sum over the lattice nodes 
implies that we must place these windows at every lattice
site, evaluate the content and add up the results of these 
evaluations. When the density profile is a
$0d$ one (as in Fig.~\ref{fig:1}a), all contributions will vanish
except those for which the window overlaps at least one node of
the $0d$ profile. In our example, this means that (\ref{eq:app1})
will return
\begin{equation}
\label{eq:spurious}
\begin{split}
\beta \mathcal{F}^{\text{ex}}_1\left[\rho_{0d}\right]=&\,
\Phi_0\left(\dnsqdp{1}{2}{3}{4}\right)
+\left[
\Phi_0\left(\dnhdp{1}{2}\right)+\Phi_0\left(\dnhdp{3}{4}\right)
\right] \\
&+\left[
\Phi_0\left(\dnvdp{1}{3}\right)+\Phi_0\left(\dnvdp{2}{4}\right)
\right]
+\sum_{i=1}^4 \Phi_0\left(\dnpdp{$i$}\right),
\end{split}
\end{equation}
where $\rho_{0d}(\bs)$ denotes a $0d$ profile of 
the form given in Fig.~\ref{fig:1}a. (Filled numbered circles
in the diagrams represent actual evalutations of the density
profile for the corresponding numbered nodes of the lattice.)

We can see that, apart from the exact value (the first term on
the r.h.s.), there appear a number of spurious contributions. Therefore
(\ref{eq:app1}) cannot be the final functional. These spurious terms
are like evaluations of $\rho_{0d}$ with non-maximal cavities,
so we will try to eliminate them by adding
new terms to (\ref{eq:app1}) corresponding to 
non-maximal cavities, with the appropriate coefficients. Since
a term like $\sum_{\bs}\Phi_0(\dnh)$ evaluated at $\rho_{0d}(\bs)$
will return $\Phi_0\left(\dnhdp{1}{2}\right)+
\Phi_0\left(\dnhdp{3}{4}\right)+
\sum_{i=1}^4 \Phi_0\left(\dnpdp{$i$}\right)$,
it seems reasonable to choose it to remove the first bracket on
the r.h.s.\ of (\ref{eq:spurious}) (and the vertical one
to remove the second bracket). Thus we use as our second ansatz
\begin{equation}
\beta \mathcal{F}^{\text{ex}}_2[\rho]=\sum_{\bs\in\ZZ}
\left[
\Phi_0\left(\dnsq\right)-\Phi_0\left(\dnh\right)-
\Phi_0\left(\dnv\right)\right].
\end{equation}
Note that in doing this, we have chosen new graphs $\mathcal{C}_k(\bs)$
and their corresponding coefficients $a_k$ in (\ref{eq:1}).
When we insert $\rho_{0d}(\bs)$ in this new functional we obtain
%\begin{equation}
$\beta \mathcal{F}^{\text{ex}}_2[\rho_{0d}]=
\Phi_0\left(\dnsqdp{1}{2}{3}{4}\right)-
\sum_{i=1}^{4} \Phi_0\left(\dnpdp{$i$}\right)$.
%\end{equation}
We have indeed removed many spurious contributions, but there
still remain some. It should now be clear that in order to remove
the latter we must add to the previous ansatz the term
$\sum_{\bs\in\ZZ}\Phi_0(\dnp)$. This way we obtain the
functional (\ref{eq:2}), which was already derived in
\cite{lafuente:2002a} by a different procedure. It is straightforward
to check that
%\begin{equation}
$\beta \mathcal{F}^{\text{ex}}[\rho_{0d}]=
\Phi_0\left(\dnsqdp{1}{2}{3}{4}\right)$,
%\end{equation}
thus proving its exact dimensional crossover.

In order to be illustrative, let us apply this procedure
again to obtain the FM functional for a different model: the
nearest-neighbor exclusion lattice gas in the triangular lattice
(hard hexagons). This example is different from the previous
one in that it has two maximal cavities: $\dntp$ and
$\dntm$. Then, the first-step functional must be
\begin{equation}
\label{eq:app1tr}
\beta \mathcal{F}^{\text{ex}}_1[\rho]=\sum_{\bs\in\ZZ}
\left[\Phi_0\left(\dntp\right)+
\Phi_0\left(\dntm\right)\right].
\end{equation}
Corresponding to the existence of two different maximal cavities
there are two different density profiles, as illustrated in
Fig.~\ref{fig:1}b. The exact $0d$ dimensional crossover must 
be satisfied for both of them. Let us start by the one
with a triangle-up shape and let us denote it $\rho_{0d}(\bs)$.
Substituting it in each of the two terms of (\ref{eq:app1tr})
we obtain (using again the window metaphor)
\begin{equation}
\begin{split}
\sum_{\bs\in\ZZ}\Phi_0\left(\dntp\right) =&\,
\Phi_0\left(\dntpdp{1}{2}{3}\right)+
2 \sum_{i=1}^3 \Phi_0\left(\dnpdp{$i$}\right), \\
\sum_{\bs\in\ZZ}\Phi_0\left(\dntm\right) =&\,
\Phi_0\left(\dntrdp{1}{3}\right)
+\Phi_0\left(\dntldp{2}{3}\right)+\Phi_0\left(\dnhdp{1}{2}\right) \\
&+\sum_{i=1}^3 \Phi_0\left(\dnpdp{$i$}\right).
\end{split}
\end{equation}
As in the previous example, to remove the ``largest'' spurious
contributions (those of the dimers) we propose
\begin{equation}
\beta \mathcal{F}^{\text{ex}}_2[\rho]=
\beta \mathcal{F}^{\text{ex}}_1[\rho]-\sum_{\bs\in\ZZ}
[ \Phi_0\left(\dntr\right)+\Phi_0\left(\dntl\right)+
\Phi_0\left(\dnh\right)].
\end{equation}
Substituting $\rho_{0d}(\bs)$ again we get
$\beta \mathcal{F}^{\text{ex}}_2[\rho_{0d}]=
\Phi_0\left(\dntpdp{1}{2}{3}\right)-
\sum_{i=1}^{3}\Phi_0\left(\dnpdp{$i$}\right)$,
so we have to add a last correction for the point-like cavities, what 
finally leads to
\begin{equation}
\label{eq:hhfinal}
\begin{split}
\beta \mathcal{F}^{\text{ex}}[\rho]=&\,\sum_{\bs\in\ZZ}
[\Phi_0\left(\dntp\right)+\Phi_0\left(\dntm\right)-
\Phi_0\left(\dntr\right) \\
&-\Phi_0\left(\dntl\right)-
\Phi_0\left(\dnh\right)+\Phi_0\left(\dnp\right)].
\end{split}
\end{equation}
This functional is exact for $\rho_{0d}(\bs)$.
We would now have to check if the same occurs for the $0d$ cavity
corresponding to the triangle-down in Fig.~\ref{fig:1}b, but
symmetry considerations immediately show that this
is the case. In general, checking dimensional crossover for a
new $0d$ cavity may lead to the appearance of additional
spurious contributions. These have to be eliminated
by adding the corresponding terms to the functional.

Finally, notice that (\ref{eq:hhfinal}) coincides with the functional
obtained in \cite{lafuente:2003p} through a completely different
(and far more involved) route.

Let us summarize the procedure to follow for an
arbitrary lattice gas with hard-core interaction.
The steps are:
%\begin{enumerate}\setlength{\itemsep}{-2pt}

(i)\ Determine the complete set of maximal $0d$ cavities of the
model. If we denote them by $\mathcal{C}_k^*$ ($k=1,\ldots,m$),
then the first-step approximation to the functional will be
[$\nbk(\bs)=\sum_{\bt\in\mathcal{C}^*_k}\rho(\bt)$]
\begin{equation}
\beta \mathcal{F}^{\text{ex}}_1[\rho]=\sum_{\bs\in\mathcal{L}}
\sum_{k=1}^m\Phi_0\left(\nbk(\bs)\right).
\label{eq:stepone}
\end{equation}

(ii)\ Select a maximal cavity $\mathcal{C}_k^*$ and let
$\rho_{0d}(\bs)$ denote a generic density profile for it.

(iii)\ Insert $\rho_{0d}$ in the current functional, 
$\beta\mathcal{F}^{\text{ex}}_i[\rho]$, 
and see which spurious contributions appear.
Identify the terms with the ``largest'' graphs
(those not contained in any other of the graphs appearing,
except $\mathcal{C}_k^*$) and pick one of them.

(iv)\ Construct the next step functional 
$\beta\mathcal{F}^{\text{ex}}_{i+1}[\rho]$ by adding to
$\beta\mathcal{F}^{\text{ex}}_i[\rho]$ a new term, with its
corresponding coefficient $a_k$, so that it
eliminates the selected spurious contribution.

(v)\ Repeat steps (iii)--(v) until no
spurious contribution remains. (Of course, one can 
exploit the symmetries of the model to resume several
steps of this process in just one, as we have done in the examples.) 

(vi)\ Repeat steps (ii)--(vi) until exhausting all maximal cavities.
%\end{enumerate} 

The functional resulting from this process will be of the form
(\ref{eq:1}) and will have, by construction, an exact $0d$
dimensional crossover. It can be proven that starting from
(\ref{eq:stepone}) there is a unique functional of the form
(\ref{eq:1}) with an exact $0d$ dimensional crossover, so
any other procedure leading to it is equally valid. (In other words,
the fact that we have chosen to remove the spurious terms
in decreasing order of ``size''
is immaterial, but in doing so we abbreviate the process.)
A sketch of the existence and uniqueness proof goes as follows
(a more detailed account will be reported in \cite{lafuente:2004}).

Let us form the set $\mathcal{P}$ with the lattice $\mathcal{L}$
and all maximal cavities $\mathcal{C}^*_k(\bs)$ ($\bs\in\mathcal{L}$,
$k=1,\dots,m$), and let us
complete it with all nonempty intersection of any number of 
maximal cavities. For any $x,y\in\mathcal{P}$, we will say that
$x\le y$ iff all nodes of $x$ are in $y$. This transforms $\mathcal{P}$
into a partially ordered set or {\em poset}. Any interval $[x,y]\equiv
\{z\in\mathcal{P}\,:\,x\le z\le y\}$ is a finite subset of
$\mathcal{P}$, so $\mathcal{P}$ is a {\em locally finite poset}.
Locally finite posets have the property \cite{stanley:1999} that 
for any mapping $f:\mathcal{P}\mapsto V$, with $V$ a vector space,
%on $\mathbb{R}$ or $\mathbb{C}$
there exists $g:\mathcal{P}\mapsto V$ such that 
%\begin{eqnarray}
%f(x) &=& \sum_{y\le x}g(y) \label{eq:inversionf} \\
%g(x) &=& \sum_{y\le x}f(y)\mu(y,x). \label{eq:inversiong}
%\end{eqnarray} 
\begin{equation}
f(x)=\sum_{y\le x}g(y), \qquad
g(x)=\sum_{y\le x}f(y)\mu(y,x).
\label{eq:inversion}
\end{equation}
The way to prove this is by inserting the second expression into the
first, what leads to 
\begin{equation}
\mu(x,x)=1, \qquad \mu(y,x)=-\sum_{y<z\le x}\mu(z,x),
\label{eq:moebius1}
\end{equation}
a recursion which defines the (integer) coefficients $\mu(y,x)$.
This scheme is referred to in the literature as a {\em M\"obius inversion},
and $\mu(y,x)$ is a M\"obius function \cite{stanley:1999}.

For the poset $\mathcal{P}$ defined above, let $\mathcal{V}$ be the space
of density functionals and take $f(x)\equiv\mathcal{F}_x^{\rm ex}[\rho]$
the (exact) excess free energy functional of a given model on the graph $x$.
Specializing (\ref{eq:inversion}) to $x=\mathcal{L}$,
\begin{equation}
\mathcal{F}_{\mathcal{L}}^{\rm ex}[\rho]=\Psi_{\mathcal{L}}[\rho]+
\sum_{x<\mathcal{L}}\big[-\mu(x,\mathcal{L})\big]\mathcal{F}_x^{\rm ex}[\rho].
\label{eq:Finversion}
\end{equation}
where $\Psi_{\mathcal{L}}[\rho]$ is an unknown functional. The sum on
the r.h.s.\ of (\ref{eq:Finversion}) only contains evaluations of 
$\mathcal{F}_x^{\rm ex}[\rho]$ for $0d$ cavities and so is an expression
similar to (\ref{eq:1}). Now let $\rho^{0d}_x(\bs)$ a generic $0d$
density profile for cavity $x$. Then,
\begin{equation}
\mathcal{F}_y^{\rm ex}[\rho^{0d}_x]=
\begin{cases}
0 & \mbox{if $x\cap y=\emptyset$,} \\
\mathcal{F}_{y\cap x}^{\rm ex}[\rho] & \mbox{otherwise.}
\end{cases}
\end{equation}
As $x\cap\mathcal{L}=x$, evaluating (\ref{eq:Finversion}) for
$\rho^{0d}_x(\bs)$ yields
\begin{equation}
\Psi_{\mathcal{L}}[\rho^{0d}_x]= 
\sum_{z\leq x} \nu(z,x) \mathcal{F}_z^{\rm ex}[\rho],\quad
\nu(z,x) = \sum_{y\cap x=z}\mu(y,\mathcal{L}),
\end{equation}
and it is a consequence of Weisner's theorem \cite{stanley:1999}
that $\nu(z,x)=0$ for any $x<\mathcal{L}$; therefore
$\Psi_{\mathcal{L}}[\rho^{0d}_x]=0$ or, in other words, the
sum on the r.h.s.\ of (\ref{eq:Finversion}) is exact for any
$0d$ cavity. 

This completes the proof that the requirement of an exact dimensional
reduction to $0d$ cavities leads to a functional of the form
(\ref{eq:1}). As to the uniqueness, it suffices to realize that
$\nu(z,x)=0$ (a necessary condition for a functional of the form
(\ref{eq:1}) to have an exact dimensional reduction to $0d$
cavities) is a particular case of the recurrence (\ref{eq:moebius1}),
whose only solution is $\mu(x,\mathcal{L})$. 

With this method one can easily recover all functionals
previously obtained in Refs.~\cite{lafuente:2002a,lafuente:2003b,
lafuente:2003p} and obtain those of virtually
any other hard-core lattice gas \cite{t-model}. 
One further striking feature of all functionals obtained in this way is
that they also have an exact dimensional crossover to one dimension,
simply because the exact one-dimensional functional is of the
form (\ref{eq:1}) \cite{lafuente:2002a,lafuente:2003p}.

Clearly the procedure presented above has no restriction in
its application other than the determination of the maximal
cavities. It can be applied to particles of any shape,
in any lattice (including regular lattices,
Bethe lattices, Husimi trees, etc.) and in any dimension.
It can even be applied to mixtures, either additive or non-additive,
provided a $0d$ cavity is properly defined as a superposition of
cavities, one for each species, such that at most one particle of
only one species can be placed in it (see \cite{lafuente:2002a}
for more details). 

The readers familiar with Kikuchi's cluster variation method
may have recognized a similarity with the procedure we have
presented here. The connection is more prominent through the
M\"obius inversion formula \cite{morita} and
will be properly discussed elsewhere \cite{lafuente:2004}.

%The accuracy of the resulting functionals when applied to
%calculate the bulk thermodynamics and correlations has
%already been discussed in detail in
%Refs.~\cite{lafuente:2002a,lafuente:2002b,lafuente:2003b,lafuente:2003p}
%for a wide set of models. In this regard, we point out that: (i) 
%the theory is very accurate at low and high densities (but, of
%course, given its mean-field nature, it is usually inaccurate
%near critical points); (ii) as it was stressed in \cite{lafuente:2003p},
%all functionals obtained in this way have consistent dimensional
%reductions; (iii) because of the latter, the functionals
%perform very well in highly inhomogeneous situations,
%and (iv) the direct correlation function obtained from this
%functionals has exactly the range of the interaction and
%recovers the Mayer function in the low density limit.

Finally, the theory can be generalized in several ways.
First of all, we have already mentioned that there
is a straightforward extension to mixtures which recovers the
functionals for mixtures already derived in \cite{lafuente:2002a,
lafuente:2003b,lafuente:2002b}. 
%The extension is valid even
%for some nonadditive mixtures (although in this case the exact 
%dimensional crossover to one dimension is lost in general;
%for exceptions see ref.~\cite{lafuente:2002a}).
A second extension is the inclusion of ``extended''
$0d$ cavities in which there can be up to $n$ particles. 
We have already checked that the
inclusion of two-particle $0d$ cavities for the Ising lattice
gas (which has repulsive and attractive interactions!)
yields the functional obtained from the 
cluster variation method at the level of the Bethe approximation
\cite{ising}, which is exact in one dimension.
Finally, there is a third extension for lattice 
gases in the presence of a porous matrix that we have
already began to explore \cite{schmidt:2003}.
Work along these lines is in progress.

%%%%%%%% Referencias %%%%%%%%%%%%%%%%%%%%%%
%See Fig.~\ref{f.1}, table~\ref{t.1} and Eq.~(\ref{e.1}).
%See also~\cite{b.a,b.b}.
%%%%%%%%%%%%%%%%%%%%%%%%%%%%%%%%%%%%%%%%

%%%%%%%% Ecuaciones %%%%%%%%%%%%%%%%%%%%
%\begin{equation}
%\label{e.1}
%0\neq1
%\end{equation}
%%%%%%%%%%%%%%%%%%%%%%%%%%%%%%%%%%%%%%%%

%%%%%%% Figuras %%%%%%%%%%%%%%%%%%%%%%%
%\begin{figure}
%\onefigure{epl-template.eps}
%\caption{Figure caption.}
%\label{f.1}
%\end{figure}
%%%%%%%%%%%%%%%%%%%%%%%%%%%%%%%%%%%%%%%

%%%%%%%% Tablas %%%%%%%%%%%%%%%%%%%%%%%
%\begin{table}
%\caption{Table caption.}
%\label{t.1}
%\begin{center}
%\begin{tabular}{lcr}
%first  & table & row\\
%second & table & row
%\end{tabular}
%\end{center}
%\end{table}
%%%%%%%%%%%%%%%%%%%%%%%%%%%%%%%%%%%%%%

\acknowledgments
We acknowledge A.~S\'anchez, C.~Rasc\'on and Y.~Mart\'{\i}\-nez-Rat\'on
for their valuable suggestions.
This work is supported by project BFM2003-0180 from 
Ministerio de Ciencia y Tecnolog\'{\i}a (Spain).

\end{document}